\documentstyle[12pt,aasms4]{article}
\begin{document}
\newcommand{\lya}{Lyman~$\alpha$}
\newcommand{\lyb}{Lyman~$\beta$}
\newcommand{\za}{$z_{\rm abs}$}
\newcommand{\ze}{$z_{\rm em}$}
\newcommand{\cmtwo}{cm$^{-2}$}
\newcommand{\nhi}{$N$(H$^0$)}
\newcommand{\nzn}{$N$(Zn$^+$)}
\newcommand{\ncr}{$N$(Cr$^+$)}
\newcommand{\degpoint}{\mbox{$^\circ\mskip-7.0mu.\,$}}
\newcommand{\halpha}{\mbox{H$\alpha$}}
\newcommand{\hbeta}{\mbox{H$\beta$}}
\newcommand{\hgamma}{\mbox{H$\gamma$}}
\newcommand{\kms}{\,km~s$^{-1}$}      
\newcommand{\minpoint}{\mbox{$'\mskip-4.7mu.\mskip0.8mu$}}
\newcommand{\mv}{\mbox{$m_{_V}$}}
\newcommand{\Mv}{\mbox{$M_{_V}$}}
\newcommand{\peryr}{\mbox{$\>\rm yr^{-1}$}}
\newcommand{\secpoint}{\mbox{$''\mskip-7.6mu.\,$}}
\newcommand{\sqdeg}{\mbox{${\rm deg}^2$}}
\newcommand{\squig}{\sim\!\!}
\newcommand{\subsun}{\mbox{$_{\twelvesy\odot}$}}
\newcommand{\et}{et al.~}

\def\ltsima{$\; \buildrel < \over \sim \;$}
\def\simlt{\lower.5ex\hbox{\ltsima}}
\def\gtsima{$\; \buildrel > \over \sim \;$}
\def\simgt{\lower.5ex\hbox{\gtsima}}
\def\arcs{$''~$}
\def\arcm{$'~$}

\title {STRONG REDSHIFT CLUSTERING OF DISTANT GALAXIES\altaffilmark{1}}
\author{\sc Judith G. Cohen\altaffilmark{2}, David W. Hogg\altaffilmark{3},
Michael A. Pahre\altaffilmark{2} and Roger Blandford\altaffilmark{3}}

\altaffiltext{1}{Based in large part on observations obtained at the
W.M. Keck Observatory, which is operated jointly by the California 
Institute of Technology and the University of California}
\altaffiltext{2}{Palomar Observatory, Mail Stop 105-24,
California Institute of Technology}
\altaffiltext{3}{Theoretical Astrophysics, California Institute of Technology,
Mail Stop 130-33, Pasadena, CA 91125}

\begin{abstract}
We present initial results from a redshift survey carried out with the
Low Resolution Imaging Spectrograph on the 10~m W. M. Keck Telescope
of a field 14.6 arc-min$^2$ in solid angle.  In the redshift
distribution of the 106 extragalactic objects in this sample we find five
strong peaks, with velocity dispersions of ${\sim}500${\kms}.  There
is evidence for a non-uniform areal density of objects in at least two
peaks.  These peaks have characteristics (velocity dispersions,
density enhancements, spacing, and spatial extent) similar to those of
nearby galaxy structures (e.g., walls and clusters), and these are
expected in a survey of this kind.  We suggest that the prominence
of these structures in our survey relative to that in other surveys
can be attributed to our $K$-selection and dense sampling.
\end{abstract}

\keywords{Cosmology: observations --- Galaxies: redshift and distances ---
Large-scale structure of Universe}

\section{INTRODUCTION}

This the first in a series of papers describing the results of a deep
survey of faint field galaxies in a single field centered at
RA(J2000)~$00^h\,53^m\,23^s\!\!.20$,
Dec.~$+12^{\circ}\,33'\,57''\!\!.5$.  The field is from the HST Medium
Deep Survey (Griffiths et al 1994), randomly selected on the basis of
high Galactic latitude ($b = -50^{\circ}$) and low reddening ($A_V =
0.13~{\rm mag}$, Burstein \& Heiles 1982).  Our redshifts were
acquired with the Low Resolution Imaging Spectrograph (Oke et al 1995)
on the 10~m W. M. Keck Telescope over a rectangular strip 2 x 7.3
arcmin$^2$ extending north-south, centered on the HST field.  Our
primary sample comprises all 155 objects with $K < 20$ mag in the 2 x
7.3 arcmin$^2$ field.  The photometry and the definition of the sample
for spectroscopic work is described in Paper II of this series, Pahre
et al (1996).  We reach higher galaxy surface densities than the
$I$-selected CFRS survey (LeF\`evre et al 1995); our survey
complements and extends to fainter objects the $K$-selected sample of
the Hawaii group (Songaila et al 1994).

Paper III in this series (Cohen et al 1996) provides a detailed
description of the redshift survey.  Of the 155 objects in the sample,
90 have spectra typical of normal galaxies, three are quasars or
broad-lined AGNs and 19 are Galactic
stars.  Of the remaining objects, 35 were observed, but no redshift
could be determined, and eight were not observed at all. The effects of
incompleteness in the sample selection and redshift identification
are discussed in Papers II and III; both are irrelevant for the
present work.  The median redshift $z$ of the 93 extragalactic objects in the
main sample is $z =$ 0.57.  An additional 13 galaxy redshifts were
determined for objects in this field which were slightly fainter than
the $K < 20$ mag limit or which lie slightly outside the spatial
boundaries of the $K$ selected sample.

\section{REDSHIFT DISTRIBUTION}

The redshift histogram is shown in Figure 1.  Of the 106 objects, 40
are in the two strongest peaks, and 64 are in the 5 strongest, i.e. have
$|z-z_p| < 0.020$ for $z_p$ = 0.392, 0.429, 0.581, 0.675, and 0.766.  In
spite of the fact that there are no clusters apparent in the images of
this field, the objects are highly clustered in redshift space.  The
overdensities in redshift space are at least a factor of 5; and more
than 60~percent of the objects in the sample lie in these structures.


\subsection{Feature significance}

We now consider the probability that these apparent features might
arise by chance out of a smooth galaxy distribution.  This
significance calculation is performed not in redshift but in the quantity
$V\equiv c\ln(1+z)$.  This coordinate is devoid of global meaning but,
granted an overall Hubble expansion, corresponds incrementally to
local velocity differences.  The galaxies are contained in an interval
$4\times10^4<V<2.4\times10^5~{\rm km\,s^{-1}}$.  As there is no
adequate {\em a priori\/} understanding of the population from which
these galaxies are drawn, the measured data set is used to derive
smoothed velocity distributions by the addition of random velocity
shifts drawn from a Gaussian distribution with width $\sigma_{V}$.
This effectively erases the obvious structure without affecting the
overall velocity distribution function.  The data set is divided into
$N_b$ uniform velocity bins and the number of galaxies $n_{0i}$ is
counted in the single bin centered on each of the candidate
associations.  This exercise is repeated using multiple realizations
of the smoothed distribution; the mean $\bar n_i$ and standard
deviation $\sigma_{ni}$ are measured for each bin.

A measure of the significance of each feature is the statistic $X_i =
|n_{0i}-\bar n_i|/\sigma_{ni}$.  To test the null hypothesis that
individual features arise by chance, we also measure the distribution
of $X_i$ in the smoothed redshift data.  Finally, as the appearance of
features like these is notoriously sensitive to the binning, we repeat
this procedure with different values of $N_b$ and, for each value of
$N_b$, a range of shifts of bin ``phase,'' comparing the maximum
values of $X_i$ measured with the maximal values produced from the
smoothed distribution.

To carry this scheme out in practice, a conservatively small smoothing
length $\sigma_{V} = 2\times10^3$~\kms\ was adopted, $N_b$ ranged from
35 to 125 (in integer increments), and for each $N_b$, 10
equally-spaced values of bin phase shift from 0 to 1 bin width were
allowed.  For each value of these parameters, 1000 realizations of the
Gaussian smoothed distribution were investigated.  The results are
given in Table 1: the observed $z_p$; the number of galaxies within
the bin coinciding with $z_p$ (this number changes slightly depending
on $N_b$; the value given is that corresponding to the $N_b$ and phase
at which $X_i$ is the largest; it is not the same as the number
assigned to each peak by the $|z-z_p|<0.020$ rule); the largest value
$X_{\rm max}$ of $X_i$ for that feature inferred from the data seen
over the full range in $N_b$ and phase; and the likelihood of the
feature arising by chance, computed as the fraction of realizations of
the smoothed distribution in which such a large value of $X_i$ was
found in any bin.  The latter quantity shows that the feature at
z=0.392 is not significant and the one at z=0.766 is only marginally so.
These results are robust to eliminating the few galaxies with
redshifts of lower precision, or the 13 not in the $K$-selected
sample.  The estimated significances increase with increasing
smoothing length $\sigma_{V}$.

\subsection{Velocity dispersions}

The radial velocity precision of our redshifts is unusually high for a
deep redshift survey.  We estimate that the uncertainty in $z$ for
those objects with with redshifts considered secure and accurate
(comprising 80 of the 106 galaxies) is $\approx 300$~\kms.  The
velocity dispersions for the five strongest peaks in the redshift
histogram are given in Table~2.  The velocity dispersions of the peaks
are only slightly smaller when the objects with low-precision
redshifts are excluded.  Both because of the biasing effect of
including outliers and the spreading of redshifts due to measurement
uncertainties, these velocity dispersions should be treated as upper
limits.

\section{ANGULAR DISTRIBUTION}

The angular distribution of the entire sample, as well as that of
galaxies with $|z - z_p| < 0.020$ for each of the five peaks, is shown
in Figure~2.


To provide a quantitative test of the uniformity of the areal
distributions, two-dimensional Kolmogorov-Smirnov tests 
(Fasano \& Franceschini
1987) were applied to the sample.  When the areal distribution of the
entire sample is compared with a uniform distribution, no evidence for
non-uniform distribution is found.  For each feature, the areal
distribution of objects in the feature (membership defined by
maximizing $X_i$ as described in the previous section) is compared
with that of all the objects not in the feature.  The 2-d K-S test $D$
values for the individual peaks are given in Table 3.  Two, at z=0.392
and z=0.581, are significantly non-uniform in areal distribution.
These results are not significantly altered if one uses only the 93
objects in the $K<20$~mag sample.  They are also robust to replacing
the distribution of feature non-members with a uniform distribution.

Interestingly, the one feature, at $z=0.392$, judged insignificant in
redshift clustering shows significant angular clustering, and the
feature, at $z=0.766$, judged marginally significant in redshift
clustering is also marginally significant in angular clustering.  For
this reason we judge both these features at least marginally
significant.

\section{DISCUSSION}

\subsection{Effects of sample definition decisions}

Although there are many other faint object redshift surveys, none have
found as much strong redshift clustering as is presented here.  We
believe that this can be attributed to differences in sample
definition.  This survey of objects goes to high galaxy number
density, $4\times 10^4$ per square degree at $K=20$, comparable to the
``Hawaii'' survey (Songaila et al, 1995) but with greater completeness at the
faint end, and deeper than the CFRS (LeF\`evre et al 1995) or the
$B$-selected LDSS--2 survey (Glazebrook et al 1995).  Infrared
selection is not subject to the same biases towards late-type spirals
and irregulars as is found in $B$-selected samples. 
Our redshifts are measured with good
precision, allowing good resolution of the peaks and their velocity
dispersions.

Finally and most importantly, this sample is not sparse-sampled, i.e.,
(almost) every object with $K<20$~mag in the field is observed, so we
do not miss structures of limited angular extent.  For very sparsely
sampled data, one obtains no statistically valid peaks at all in the
redshift distribution, c.f.\ the CFRS (Lilly et al 1995).  As one
samples less sparsely, ``walls'' with substantial velocity
dispersions, as in the ESO survey (Bellanger \& de Lapparent 1995),
appear.  Suggestions similar to these regarding the effect of
different sampling schemes have been offered by de Lapparent et al
(1991) and by Ramella et al (1992) to explain why some redshift
surveys did not see structures such as the ``Great Wall'' while
others, e.g., the CFA survey (de Lapparent et al 1986), did.

\subsection{Structure morphology}

In general, imaging surveys for galaxy structures find clusters and
filaments because these appear as angular patches of higher projected
number density, but do not find walls because the projected number
density is relatively invariant to collapse from three-dimensional
volumes into two-dimensional walls.  On the other hand, redshift
surveys like this one, in small angular areas, are unlikely to hit
compact clusters and filaments, but should pierce any walls.
Because the structures in the present sample
are seen in the redshift distribution but not
in the imaging, there is some evidence that they are wall-like.

The derived velocity dispersions $\sigma_v$ are small and reminiscent
of those of sparse clusters of galaxies.  Zabludoff et al (1990) found
a median $\sigma_v =740$~{\kms} for 69 nearby, moderately rich Abell
clusters.  Our observed $\sigma_v$ are much smaller than that found
for superclusters; for example, Postman, Geller, \& Huchra (1988)
found $\sigma_v =1300$~{\kms} for primary members of the seven Abell
clusters in the Corona Borealis supercluster, while Small (1996) found
1800~{\kms} for the entire supercluster.  If the cluster Cor Bor
itself is excluded, then the value drops to $<1000$\kms (Postman,
Huchra, \& Geller 1992; Zucca et al 1993).  Small (1996) also observed
many individual lines of sight through the supercluster and found
$400<\sigma_v <1200$~\kms; these lines of sight are comparable in
extent to our field, while our derived $\sigma_v$ are among lowest
values measured along individual Cor Bor lines of sight.  The velocity
dispersion for the Great Wall, however, is 230~{\kms} measured
over 3$^\circ$ cells (Ramella et al 1992), while that for groups
in the SSRS2 (DaCosta et al 1994) is $\sim 200$~\kms.  Our $\sigma_v$
are thus comparable to Great-Wall-size structures, poor clusters of
galaxies, or lines of sight through superclusters.  The separations
between structures are also comparable: the comoving distances of the
five strongest redshift peaks are 915, 981, 1228, 1364, and
1485$h^{-1}$~Mpc, while the distance between the Cor Bor supercluster
and the one immediately behind it at $z \approx$ 0.12 (Goia et al
1982, Sarazin et al 1982, and Small 1996) is 130$h^{-1}$~Mpc.  In this
regard our redshift distribution is also reminiscent of that found by
Broadhurst et al (1990), although the features in ours are not
``periodic.''

At $z$ = 0.6, our field has a size of $1.7\times 0.5\,h^{-2}$~Mpc$^2$.
We take the number density of clusters in the universe to be of order of a few
times $10^{-5}\,h^3~{\rm Mpc^{-3}}$, as reported for Shectman (1985)
clusters (Bahcall, 1988) and $R\geq 1$ Abell galaxy clusters (Postman
et al 1996).  The fact that we observe 5 clusters along a randomly
selected line of sight implies a typical cluster radius of between $2$
and $5\,h^{-1}~{\rm Mpc}$, depending on the exact number density and the
world model.  These lengths are on the order of, if a bit higher than,
locally measured cluster radii, and are consistent with the marginally
non-uniform angular distributions we find above.  That these
structures might be sparse clusters is also consistent with their
small measured velocity dispersions.

If we ``run the clock backward'' on structure formation, it seems
unlikely that the huge walls and rich clusters seen locally would
completely disperse into a uniform galaxy distribution by redshift of
$1$.  It would be surprising if the high-redshift counterparts of
local walls were {\em not\/} found in deep redshift surveys. (Rich
clusters, on the other hand, only contain a small fraction of all
galaxies and are much less likely to be encountered).  Many
groups have found similar or related structures.  Bellanger \& de
Lapparent (1995), presenting the first results of the ESO-Sculptor
Faint Galaxy Redshift Survey for galaxies with $R<20.5$~mag over
0.28~deg$^2$, have found what they call ``walls'' analogous to the Great 
Wall seen in the local universe.  The galaxy distribution they see is
strongly clustered in the line of sight, consisting of walls and
voids, although their redshift histogram does not show structures as
strongly peaked in redshift space as those presented here (for the
reasons given above).  A structure, interpreted as a normal dense
galaxy cluster at high redshift, was found by LeF\`evre et al (1994),
consisting of a group of 12 galaxies around a quasar at $z=0.98$, with
velocity dispersion $\sigma_v=955$~{\kms} and transverse structure on
a characteristic scale $2\,h^{-1}$~Mpc.  Of course an observation of
transverse structure is not an argument against wall morphology
because if a sheet becomes self gravitating it must break up into
structures that as wide as the wall is thick (Ostriker \& Cowie 1981).
Hutchings et al (1995) have detected compact groups of galaxies with a
radius of $\sim 1$~arcmin (0.25$h^{-1}$ Mpc) around 14 QSOs with
$z\sim 1.1$.  Ellingson et al (1991) (see also Ellingson \& Yee 1994)
find galaxy clusters around QSOs with $\sigma_v\sim 400$--$500$~\kms.
Studies of quasar absorption lines in QSO pairs and in individual
objects also suggest the existence of superclusters at $z\approx 2.5$
(Dinshaw \& Impey 1996).

\subsection{Critical future observations}

Given these observations and our interpretation, it is possible to
predict the outcome of future observations which are crucial in
determining the statistical, morphological, and physical properties of
these structures.  If they are wall-like, then we expect to see
coherence as we extend the survey to adjacent fields.  Even if the
clumps are sparse clusters, local observations of Cor Bor and the
Great Wall suggest that the clumps will group into large
two-dimensional sheets.  Redshift surveys in adjacent fields are
essential to answering these questions of morphology.

There are many redshift surveys undertaken by
different groups, and it is important to demonstrate that the
differences among these surveys in prominence  
of structures in the $z$ distribution 
is
indeed attributable to differences in sample definition.  It is
imperative that the objects in this survey (and adjacent fields) be
reselected with photometric and sparse-sampling criteria matched to
other surveys; comparison could then be used to confirm that the
structures are common and this field is not ``special.''  Also, the
morphology-density relation (Dressler 1980)
suggests that changing to bluer selection
bands should reduce the percentage of galaxies lying inside the
structures.

Finally, similar samples in widely separated fields will provide us
with statistics for these structures, such as typical separations,
velocity dispersions, filling factor, etc.  It is possible that some
of the much-discussed field-to-field variations found in galaxy counts,
angular correlation functions, and redshift distributions (e.g., Koo
\& Kron, 1992) can be attributed to the field-to-field variations in
the walls along the line of sight.

If, as we strongly suspect they will, further observations do
substantiate the view that roughly half the old galaxies are localized
in walls, then a major challenge will be to to determine whether the
galaxies in these walls have virialized.  At present there appear to
be no good observational arguments against this view.

\acknowledgements
We are grateful to George Djorgovski, Keith Matthews, Gerry
Neugebauer, Tom Soifer and Jim Westphal for helpful conversations and
to Todd Small and Wal Sargent for permission to use their data on Cor
Bor prior to publication.  The entire Keck user community owes a huge
debt to Bev Oke, Jerry Nelson, Gerry Smith, and many other people who
have worked to make the Keck Telescope a reality.  We are grateful to
the W. M. Keck Foundation, and particularly its president, Howard
Keck, for the vision to fund the construction of the W. M. Keck
Observatory.

\newpage

\begin{deluxetable}{cccc}
\tablewidth{0pc}
\scriptsize
\tablecaption{Statistical Parameters for Redshift Peaks}
\tablehead{
\colhead{$z_p$} & \colhead{No. Galaxies at $N_b$(max)} & 
\colhead{$X_{max}$} & Likelihood}
\startdata
 0.392  &  9 &  3.4 & 0.5   \nl
 0.429  & 12 &  6.4 & 0.005 \nl
 0.581  & 20 & 12.4 & 0     \nl
 0.675  &  8 &  5.3 & 0.02  \nl
 0.766  &  6 &  5.0 & 0.2   \nl
\enddata
\end{deluxetable}

\newpage
\begin{deluxetable}{cccc}
\tablewidth{0pc}
\scriptsize
\tablecaption{Velocity Dispersions in Redshift Peaks}
\tablehead{
\colhead{$z_p$} & \colhead{$N$\tablenotemark{a}} & 
\colhead{$\sigma_v(N)$}
 & \colhead{$\sigma_v(N-1)$\tablenotemark{b}} \nl
\colhead{~} & \colhead{~} &  \colhead{(\kms)} & \colhead{(\kms)}} 
\startdata
 0.392  &  9 &  585 &  465 \nl
 0.429  & 17 &  685 &  615 \nl
 0.581  & 23 &  610 &  410 \nl
 0.675  &  8 &  440 &  405 \nl
 0.766  &  7 &  975 &  670 \nl
\enddata
\tablenotetext{a}{Number of galaxies with
$z$ within 0.02 of $z_p$.}
\tablenotetext{b}{Omitting the most discrepant galaxy, velocity
dispersion as derived from the remaining $N-1$.}
\end{deluxetable}

\newpage
\begin{deluxetable}{ccc}
\tablewidth{0pc}
\scriptsize
\tablecaption{Results of 2D K-S Tests for Uniform Spatial Distribution}
\tablehead{
\colhead{$z_p$} & \colhead{No. Galaxies at $N_b$(max)} & 
\colhead{2D K-S Statistic $D$}}
\startdata
 0.392  &  9 &  0.01 \nl
 0.429  & 12 &  0.2 \nl
 0.581  & 20 &  0.01 \nl
 0.675  &  8 &  0.4 \nl
 0.766  &  6 &  0.1 \nl
\enddata
\end{deluxetable}

\clearpage 
\figcaption[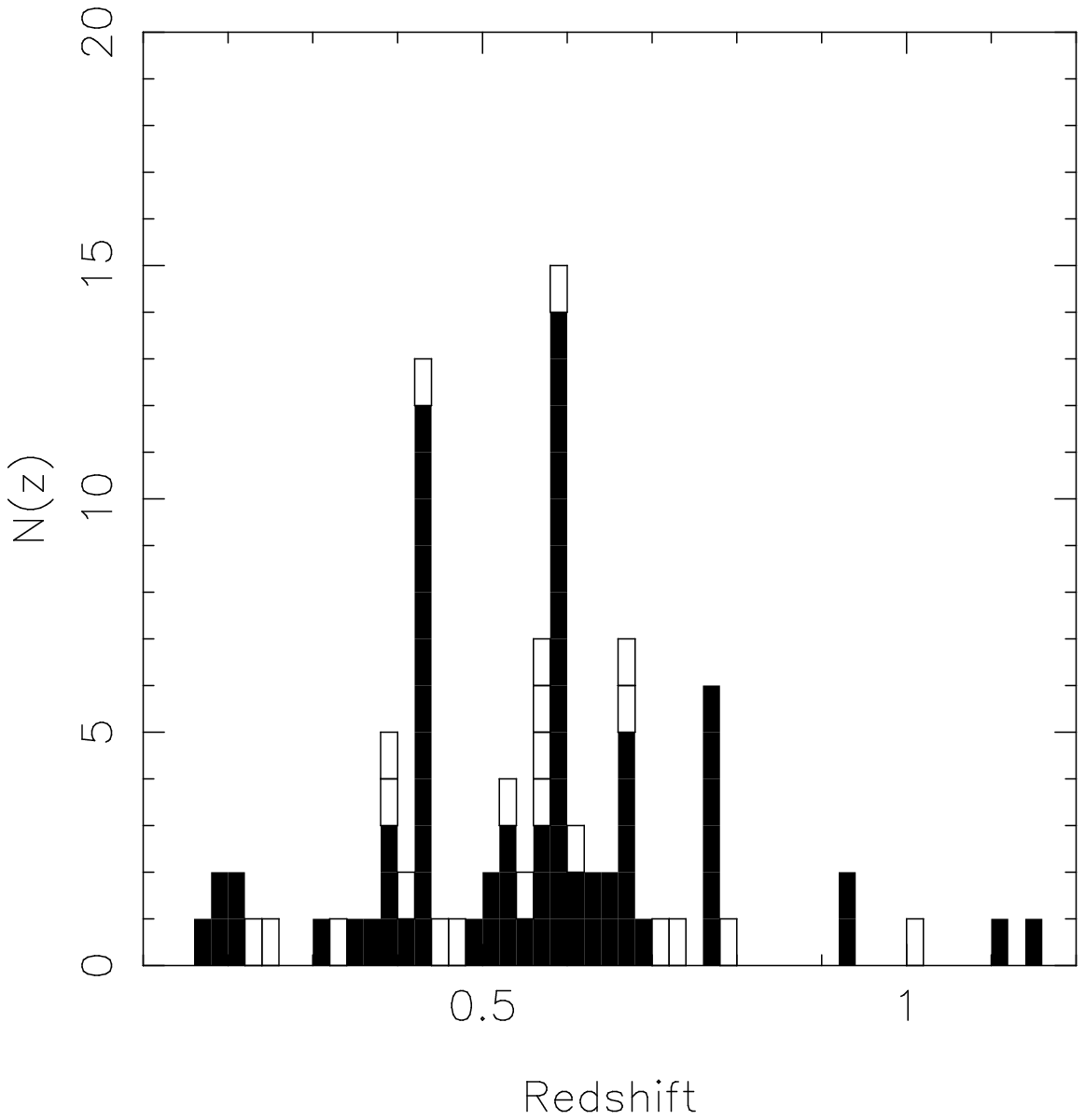]
{The redshift histogram for the galaxies in our survey.
The solid fill denotes galaxies whose redshifts are considered secure,
while the open fill denotes galaxies whose redshifts are of lower
precision.\label{fig1}}
\figcaption[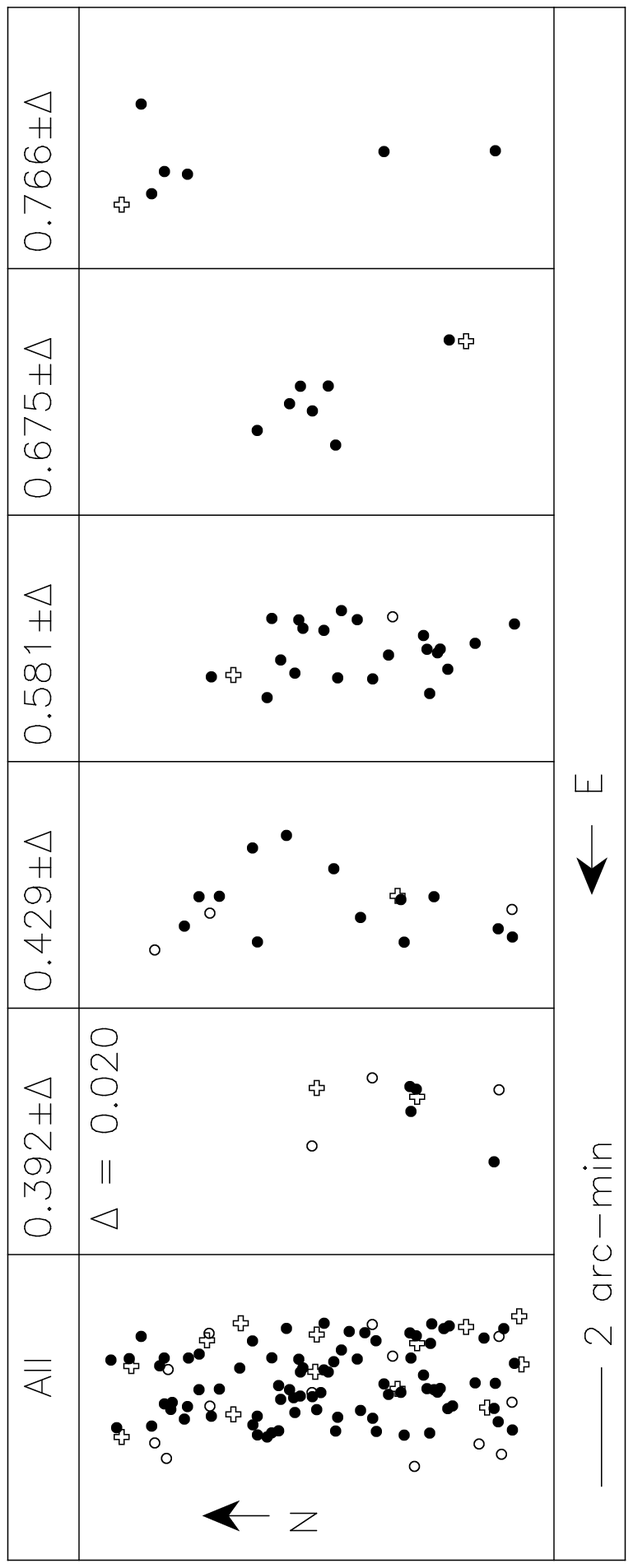]
{The distribution of our sample of galaxies projected
onto the sky is shown. The first panel shows the entire sample.
This is followed by the spatial distribution of the galaxies in each of the
5 strongest peaks in the redshift histogram.  
The galaxies not included in the $K <$ 20 mag
sample are shown as open circles, while those with uncertain $z$
are shown as open crosses.\label{fig2}}

\clearpage
\begin{figure}
\plotone{figure1.ps}
\end{figure}
\clearpage
\epsscale{0.9}
\begin{figure}
\plotone{figure2.ps}
\end{figure}
\end{document}